\documentclass[aps,prl,reprint,superscriptaddress,showpacs]{revtex4-1}
\usepackage{graphicx,amsmath,amssymb,ulem}
\usepackage{verbatim}

\usepackage{ragged2e}

\usepackage{color}

\begin{document}
\title{Separable Schmidt modes of a non-separable state}

\author{A. Avella}
\affiliation{INRIM, Strada delle Cacce 91, Torino 10135, Italy}
\affiliation{Dipartimento di Fisica, Universit\`a degli Studi di Torino, via P. Giuria 1, 10125 Torino, Italy}

\author{M. Gramegna}
\affiliation{INRIM, Strada delle Cacce 91, Torino 10135, Italy}

\author{A. Shurupov}
\affiliation{INRIM, Strada delle Cacce 91, Torino 10135, Italy}

\author{G. Brida}
\affiliation{INRIM, Strada delle Cacce 91, Torino 10135, Italy}

\author{M. Chekhova}
\affiliation{Max-Planck Institute for the Science of Light, G.-Scharowsky Str 1/Bldg 24, 91058, Erlangen, Germany}
\affiliation{M. V. Lomonosov Moscow State University, 119992 GSP-2, Moscow, Russia}

\author{M. Genovese}
\affiliation{INRIM, Strada delle Cacce 91, Torino 10135, Italy}

\date{\today}


\begin{abstract} Two-photon states entangled in continuous variables such as wavevector or frequency represent a powerful resource for quantum information protocols in higher-dimensional Hilbert spaces. At the same time, there is a problem of addressing separately the corresponding Schmidt modes. We propose a method of engineering two-photon spectral amplitude in such a way that it contains several non-overlapping Schmidt modes, each of which can be filtered losslessly. The method is based on spontaneous parametric down-conversion (SPDC) pumped by radiation with a comb-like spectrum. There are many ways of producing such a spectrum; here we consider the simplest one, namely passing the pump beam through a Fabry-Perot interferometer. For the two-photon spectral amplitude (TPSA) to consist of non-overlapping Schmidt modes, the crystal dispersion dependence, the length of the crystal, the Fabry-Perot free spectral range and its finesse should satisfy certain conditions. We experimentally demonstrate the control of TPSA through these parameters. We also discuss a possibility to realize a similar situation using cavity-based SPDC.
\end{abstract}
\maketitle

\textit{Introduction.} Entanglement of two-photon states (biphotons) in continuous variables such as frequency or wavevector suggests the use of biphotons as a quantum-information
resource in higher-dimensional Hilbert spaces~\cite{Law,ivano}. The dimensionality of the Hilbert space is determined in this case by the Schmidt number,
 the effective number of Schmidt modes, which can reach several hundred~\cite{Kulik_Fedorov,TPSA6,Gatti}.
But in order to realize any protocols with multimode states, it is necessary to address single Schmidt modes separately.
For wavevector variables, any single Schmidt mode can be filtered out using a single-mode fibre and a spatial light modulator~\cite{Straupe}.
This filtering, in principle, can be lossless, which is crucial for experiments with twin-beam squeezing~\cite{IvanoPRL,IvanoNature, mb1,mb2,mb3,Agafonov}.
For frequency variables, it is far more difficult to losslessly select a single Schmidt mode.
Attempts are being made in homodyne detection~\cite{Fabre}; in direct detection experiments the procedure is more difficult.
For instance, methods based on nonlinear frequency conversion are proposed, technically complicated to realize with high efficiency~\cite{Eckstein}. Here we propose and demonstrate a method of engineering a two-photon state in such a way that it contains \textit{non-overlapping} Schmidt modes, each of which can be filtered losslessly using a spectral device.

\textit{Frequency entanglement of biphotons.}
A two-photon state generated via SPDC  can be written in the form
	\begin{equation}
	\label{state}
\left| \psi \right\rangle \propto \int \int d\omega_{i} d\omega_{s} F(\omega_{i},\omega_{s}) \hat{a}^\dagger_{i}(\omega_{i}) \hat{a}^\dagger_{s}(\omega_{s}) \left| vac \right\rangle,
	\end{equation}
where $ \hat{a}^\dagger_{s},\hat{a}^\dagger_{i} $ are the creation operators of the signal and idler photons, and $F(\omega_{s},\omega_{i})$
represents the two-photon spectral amplitude (TPSA)~\cite{TPSA1}.
The TPSA fully characterizes the spectral properties of a biphoton  state and its physical meaning is the joint spectral probability amplitude of the down-converted photons in signal and idler modes with frequencies $\omega_{s}$ and $\omega_{i}$ respectively.
TPSA is used to determine the degree of frequency entanglement \cite{Law,TPSA6,TPSA3,TPSA4,TPSA5} and plays a central role in the heralded generation of pure single-photon states~\cite{Mosley}.

The TPSA depends on both the pump spectrum $F_p(\omega)$ and the phase matching in the nonlinear crystal,
\begin{equation}
	\label{TPSA}
F(\omega_{s},\omega_{i}) = e^{i\Delta k_zL/2}F_p(\omega_{s}+\omega_{i}-\omega_{p})\,\hbox{sinc}(\Delta k_zL/2),
	\end{equation}
where $\omega_p$, $\omega_{s},\omega_{i}$ are the pump, signal, and idler frequencies, $\hbox{sinc}(x)\equiv\sin x/x$, $\Delta k_z\equiv\Delta k_z(\omega_{s},\omega_{i})$  is the longitudinal mismatch, and $L$ the crystal length.

If the TPSA is not factorable, $F(\omega_{s},\omega_{i})\ne F_s(\omega_s)F_i(\omega_i)$, then the state (\ref{state}) is entangled. Nevertheless, it  can always be written as a sum of factorable states (the Schmidt decomposition),
\begin{eqnarray}
	\label{SchmidtTEO}
\left| \psi \right\rangle =\sum_{n=0}^\infty\sqrt{\lambda_n}\int d\omega_{s} f^s_n(\omega_{s}) \hat{a}^\dagger_{s}(\omega_{s})\left| vac \right\rangle\nonumber\\
\times\int d\omega_{i} f^i_n(\omega_{i}) \hat{a}^\dagger_{i}(\omega_{i}) \left| vac \right\rangle,
	\end{eqnarray}
where $f^{s,i}_n(\omega)$ are the Schmidt modes for the signal and idler photons and $\lambda_n$ are the Schmidt coefficients, $\sum_n\lambda_n=1$. The Schmidt number $K\equiv[\sum_n\lambda_n^2]^{-1}$ represents a measure of entanglement \cite{z}.

\begin{figure}[h!]
\includegraphics[width=0.9\columnwidth]{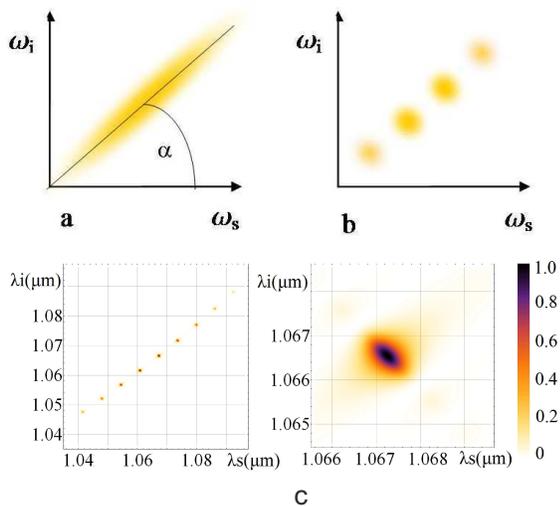}
\caption{a,b: A simplified picture of the TPSA formed by a short pump pulse (a) and the same pulse after passing through an FP cavity (b). c: Squared modulus of the TPSA calculated for a 20 mm KDP crystal pumped by 25 fs pulses at 532 nm transmitted through a $50 \mu$m FP cavity (left) and the zoom into a single maximum (right).} \label{Schmidt}
\end{figure}

\textit{Separable Schmidt modes.} In most cases, filtering out a single Schmidt mode $f_n^{s,i}$ is challenging. However, the task becomes simple if the modes of different orders do not overlap in frequency, $\forall\omega\,\,f_n^{s}(\omega)f_m^{s}(\omega)\sim\delta_{nm}$. (Due to the orthogonality of the Schmidt modes a weaker condition is always fulfilled, $\int d\omega f_n^{s}(\omega)f_m^{s*}(\omega)=\delta_{nm}$). Then, with the help of an appropriate spectral device, the simplest one being a prism followed by a slit, any of the signal Schmidt modes $f_n^{s}(\omega)$ can be filtered, in principle, losslessly, and similarly for the idler modes.
As an example, consider SPDC from a pump with a comb-like spectrum. Such a pump can be obtained using standard pulse shaping methods; as a proof-of-principle demonstration we will consider a simple one, based on passing a laser beam through a Fabry-Perot (FP) cavity (Fig.~\ref{Schmidt}).
If the phase matching leads to a TPSA stretched in the $\omega_s+\omega_i$ direction (a), the introduction of an FP with an appropriate free spectral range will result in a TPSA given by separate maxima (b). If these maxima give non-overlapping projections on both horizontal and vertical axes, each of them represents a product of Schmidt modes $f_n^{s}(\omega_s)f_n^{i}(\omega_i)$~\cite{Supp}.

In order to realize this situation in experiment, one can notice that the tilt of the TPSA (Fig.\ref{Schmidt}a) can be changed by using different phase matching conditions.
Indeed, the tilt $\alpha$ is given by $\tan\alpha=-\gamma_s/\gamma_i$, where $\gamma_{s,i}\equiv u_p^{-1}-u_{s,i}^{-1}$, $u_{p,s,i}$ being the group velocities of the pump,
signal, and idler photons~\cite{Chekhova}. In most cases, the tilt is negative, but for frequency-degenerate type-II phasematching in KDP crystal from a $415$ nm pump, $\gamma_s=0$, hence a zero tilt can be realized~\cite{Mosley}. The possibility of a positive tilt,
first discussed in Ref.~\cite{Giovannetti}, is realized in KDP at higher pumping wavelengths. The tilt is $45^\circ$ if KDP is pumped at 532 nm (Fig.\ref{Schmidt}c). The shape of a single TPSA maximum in the presence of a comb-like
pump spectrum can then be changed by changing the length of the crystal and the width of a single maximum in the `comb'.

\textit{Gaussian model.} Consider now a single 'spot' of the TPSA distribution in Fig.~\ref{Schmidt}. Its shape can be obtained from Eq.~(\ref{TPSA}) by assuming that the pump spectrum
is given by a single peak of the `comb'. For simplicity, let us first describe the shape of this peak as a Gaussian and replace the $\mathrm{sinc}$ function by a Gaussian function as well.
Then, the cross-section of the TPSA for a single `spot' in Fig.~\ref{Schmidt}b is represented by an ellipse, which will be oriented horizontally or vertically if~\cite{Supp}
 \begin{equation}	
 \label{condition}
\sin(2\alpha)=\sigma_c^2/\sigma_p^2,
	\end{equation}
where $\sigma_c$ is related to the crystal length L, and $\sigma_p$ is given by the width of a single peak in the pump spectrum.
This way of obtaining a single-mode TPSA is similar to the one used in Ref.~\cite{EcksteinPRL}. Clearly, condition (\ref{condition}) can be only satisfied for $\alpha>0$.
The required value of $\alpha$ is the smaller, the longer is the crystal and the broader is the width of a single `comb' maximum.  This provides additional  possibilities to engineer the state. For the maxima to overlap neither in the signal frequency nor in the idler one, they should be sufficiently well separated. This imposes additional requirement on the distance between the `comb' maxima $\Delta\omega$. For $\alpha<45^\circ,$ the condition is $\sin^4\alpha/\sigma_c^2+\sin^2\alpha/2\sigma_p^2>>1/(\Delta\omega)^2$~\cite{Supp} and it can be satisfied for a sufficiently large $\Delta\omega$.

\begin{figure}[h!]
\includegraphics[width=0.9\columnwidth]{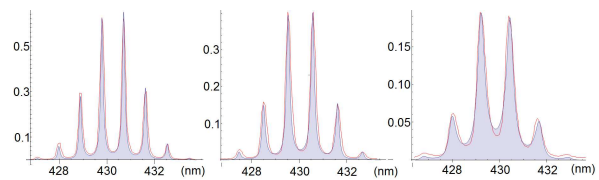}
\caption{Measured (red line) and calculated (filled contour) pump spectrum with the central wavelength $430$ nm and the width $2.8$ nm, transmitted through a FP interferometer normal to the beam (left), tilted by $20^\circ$ (center) and  $45^\circ$ (right).} \label{FP}
\end{figure}

\textit{Numerical calculation.} An exact numerical calculation of the TPSA has been performed for the case of a type-II KDP crystal pumped by $160$ fs pulses with
the central wavelength varying from $370$ nm to $450$ nm, transmitted through an FP interferometer with the thickness of $100\,\mu$m. Typical pump spectrum is shown in Fig.~\ref{FP}. By tilting the FP (middle and right panels), one can change the finesse and hence the width of a single maximum.

\begin{figure}[h!]
\includegraphics[width=0.95\columnwidth]{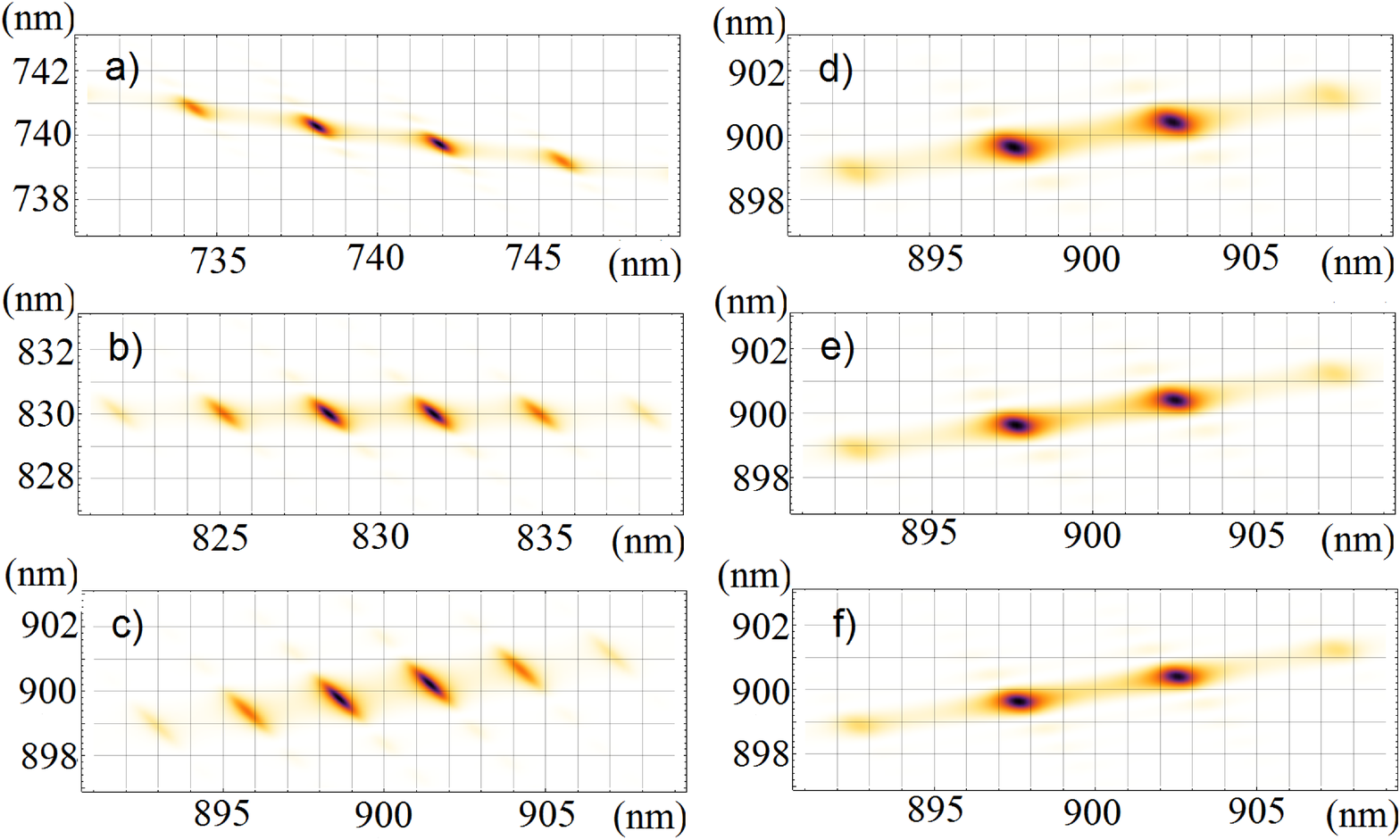}
\caption{Calculated squared modulus of TPSA for type-II SPDC in a KDP crystal pumped by $160$ fs pulses passed through a 100$\mu$m FP interferometer. Left: 15 mm crystal, the pump central wavelength is 370 nm (a), 415 nm (b) and 450 nm (c) and the FP is at normal incidence. Right: the pump central wavelength is $450$ nm, the FP is tilted by 45 degrees and the crystal thickness is $13$ mm (d), $15$ mm (e), and $17$ mm (f).} \label{kdp_calc}
\end{figure}

Calculated TPSA shapes for different pump wavelengths are shown in Fig.~\ref{kdp_calc} a-c.
One can see that the tilt of the TPSA changes, being negative for wavelengths below $415$ nm and positive otherwise. The FP orientation is assumed to be normal and the crystal length
is $15$ mm. One can observe that even in panel c, separate `spots' of the TPSA do not represent single-mode states as they are stretched in a tilted direction.
The only ways to make them single-mode are either to increase the crystal length or to broaden the FP transmission peaks.
This possibility is demonstrated in Fig.~\ref{kdp_calc} d-f showing the case of the FP tilted by $45^\circ$ and the crystal length $13$ mm
(d), $15$ mm (e), and $17$ mm (f). The TPSA shown in Fig.~\ref{kdp_calc}e has the desired feature: each separate maximum represents a single-mode state. The small overlap of the neighboring maxima, partly caused by the background of the FP transmission spectrum, can be eliminated by simultaneously increasing the FP finesse and the crystal length, which was technically impossible in our experiment. In the case of pumping at $532$ nm this problem does not arise (Fig.~\ref{Schmidt}c) as the `spots' are well separated in both dimensions.

For an isolated single `spot' of the TPSA in Fig.~\ref{Schmidt}c, exact numerical calculation of the Schmidt number gives $K=1.23$~\cite{Supp}. The deviation from the unity is caused by the pump Lorentzian shape and the side lobes of the `sinc' function. Shaping the pump spectrum as a comb of Gaussian peaks would give $K=1.06$.

\begin{figure}[h!]
\includegraphics[width=0.90 \columnwidth]{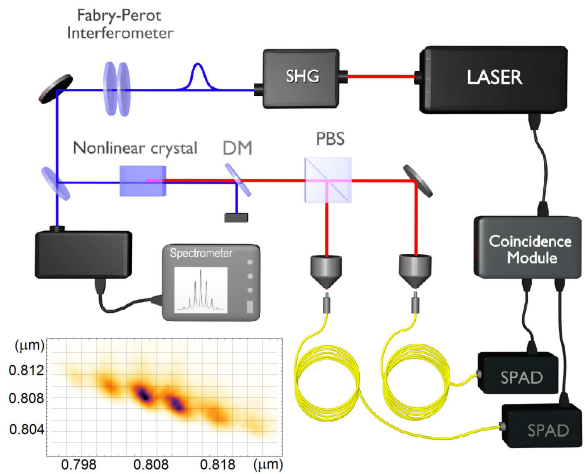}
\caption[blablabla]{ (Color online) The experimental setup:
frequency-doubled pulses from mode-locked Ti-Sa laser are fed into an
FP interferometer shaping the spectrum as shown in Fig.\ref{FP}. The beam pumps a type-II nonlinear crystal (BBO or KDP). A dichroic mirror
(DM) removes the residual pump beam. Photons of the same pair are separated by a polarizing beam splitter and fed in two identical fibers of $1$ km length. Finally, photons are
detected by two SPADs and coincidence electronics is used to measure their arrival times
with respect to the trigger from the laser. Inset: the measured distribution, recalculated into wavelengths, for a $5$ mm BBO crystal.}
\label{se3tup}
\end{figure}

\textit{Experimental setup.} In the preparation part of our setup (Fig. \ref{se3tup}) we use a Ti-Sapphire mode-locked laser, with a pulse duration of $160$ fs, a Gaussian spectrum with the central wavelength tunable around $800$ nm with  FWHM bandwidth of $10$ nm, frequency-doubled to get a FWHM bandwidth of $2.8$ nm. Into the frequency-doubled beam we put an FP cavity with air spacing of $100$ $\mu$m in order to shape the spectrum like in Fig.\ref{FP}. The beam is then focused, with an $f = 1$m lens, into a single $5$ mm BBO crystal or a pair of $5$ mm KDP crystals cut for collinear degenerate type-II phasematching. In order to reduce the effect of transverse walk-off, the crystals have optic axes tilted symmetrically with respect to the pump direction. After the crystals, the pump is eliminated using a dichroic mirror and a red-glass filter.

In the registration part of the setup, the TPSA is measured using the effect of frequency-to-time Fourier transformation in the course of two-photon light propagation through a dispersive medium. This effect, first studied for cw biphotons~\cite{masha1,Valencia,masha2}, was later applied to the spectroscopy of single photons~\cite{baek} and next to the measurement of TPSA for femtosecond-pulsed biphotons~\cite{Silberhorn,masha3,EcksteinPRL}.
The signal and idler photons are sent through different optical fibres, and then their arrival times with respect to the pump pulse are analyzed~\cite{EcksteinPRL}. After a fibre of length $l$ the joint probability distribution amplitude of the arrival times for signal and idler photons $\widetilde{F}(t_s, t_i)$ takes the shape of the TPSA $F(\Omega_s,\Omega_i)$, with the frequency arguments rescaled~\cite{masha3},
\begin{eqnarray}
	\label{TPTA}
\widetilde{F}(t_s, t_i) \propto \exp\{-i\frac{t_i^2}{2k''_i l}  -i\frac{t_s^2}{2 k''_s l} \}  F(\Omega_s,\Omega_i),\nonumber\\ \quad
	\Omega_s = t_s/k''_s l, \quad \Omega_i = t_i/k''_i l,
\end{eqnarray}
 the tilde denoting the Fourier transformation
and $k''_{s,i}$ given by the group-velocity dispersion (GVD) of the fibre.

In our measurement setup, the two photons of the same pair are split on a polarizing beam splitter and fed in two identical Nufern 780-HP fibres of $1$km length, with the GVD changing from $-120$ ps/nm/km at $800$ nm to $-90$ ps/nm/km at $900$ nm. At the fibre outputs, two silicon-based single photon avalanche diodes (SPADs) with $50$ ps time jitter, connected to a three-channel time-to-digital converter, measure the distribution of the arrival times of the signal and idler photons with respect to the trigger signal of the pump pulse provided by a fast photodiode inside the laser housing.
The measurement yields a histogram proportional to squared TPSA, with the resolution of $1.5$ nm. As an example, the inset to Fig.~\ref{se3tup} shows the distribution obtained for a $5$ mm BBO crystal pumped by a $404$ nm pump. The arrival times are recalculated into wavelengths according to Eqs.~(\ref{TPTA}). Separate maxima of TPSA are clearly seen but they do not represent single-mode states, due to the overall negative tilt of the TPSA. In order to obtain the positive tilt, we made the measurements with KDP crystals and the pump wavelength varying within the range $395-440$ nm.

\begin{figure}[h!]
\includegraphics[width=0.95\columnwidth]{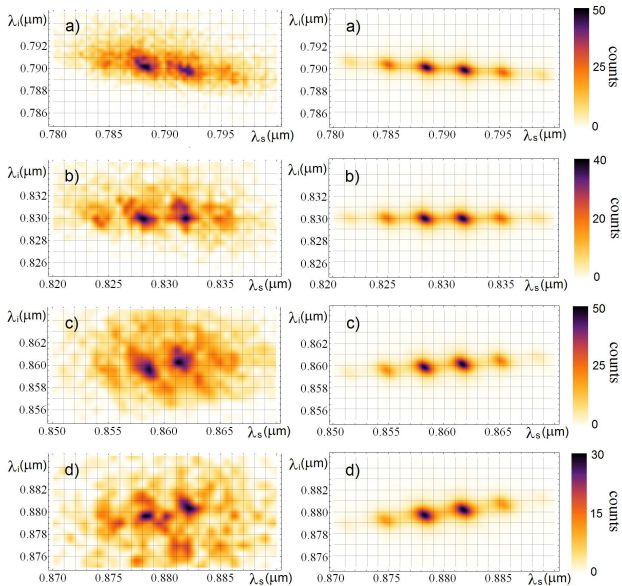}
\caption{Left: measured squared modulus of TPSA (in counts) for type-II KDP crystal of total length $10$mm pumped at (a) $395$ nm, (b) $415$ nm, (c) $430$ nm, (d) $440$ nm. The FP interferometer is tilted at $45^\circ$. Right: the corresponding theoretical distributions with $1.5$ nm resolution taken into account.} \label{exp}
\end{figure}
\textit{The results of the measurement} with the KDP crystals are presented in Fig. \ref{exp} together with the distributions calculated with an account for the finite time resolution. As the pump wavelength changes from $395$ nm to $440$ nm, the TPSA tilt changes from negative to positive. One clearly sees the `spot' structure caused by the FP transmission maxima. The number of distinct transmission maxima is reduced due to the FP tilt (see Fig.~\ref{FP}). Despite the experimental non-idealities (spreading of each `spot' due to the detectors' jitter, high background caused by low FP finesse and the reflections at the crystal surfaces, and large statistical error caused by weak signals), several important observations can be made. The distribution with the negative TPSA tilt (Fig.~\ref{exp}a), similarly to the inset to Fig.~\ref{se3tup}, shows the multi-mode structure of separate spots. The distribution in Fig.~\ref{exp}b shows nearly single-mode structure of separate spots but in this case, the whole TPSA is factorable and thus represents a single-mode state. Finally, in the positive-tilt distributions (Fig.~\ref{exp}c,d) each spot, according to the theoretical predictions, has single-mode structure and, at the same time, the whole TPSA is not factorable.

\textit{Cavity SPDC}. A TPSA with non-overlapping Schmidt modes can be realized in a different way, in which not the pump has a `comb-like' spectrum but the signal and idler photons. This is the case in cavity-enhanced SPDC~\cite{Ou,cavity} with type-I phase matching, a negative TPSA tilt, and the cavity resonant for signal/idler radiation. The cavity will then select narrowband Lorentzian maxima at the same signal and idler frequencies. Provided that the pump spectrum is considerably broader than these maxima (but still much narrower than the cavity FSP), the TPSA will consist of single-mode maxima displaced in the direction $\omega_s+\omega_i=\hbox{const}$~\cite{Thew}. For the experimental parameters of Ref.~\cite{Ou}, this situation will be realized for a few nanosecond pump pulses and will result in the maxima of about 100 MHz width. Note that a spectrum with a similar structure has been considered in Ref.~\cite{Lu}; such states were called mode-locked two-photon states. The issue of separable Schmidt modes, however, was not discussed. In many current experiments, a single maximum can be filtered out~\cite{Mitchell}, and it can be a Schmidt mode if more broadband pumping is used.

\textit{Conclusion}. We have shown that it is possible to prepare a biphoton state with multimode frequency-temporal structure containing non-overlapping Schmidt modes. Such a state, provided that the number of modes is large, can be used for higher-dimensional encoding of quantum information, with the possibility to address separate modes. Although in our proof-of-principle experiment there are only few Schmidt modes, a highly multimode state can be obtained by shaping the pump pulse using more advanced techniques. A similar method of creating non-overlapping Schmidt modes can be applied to the spatial TPSA; then the pump spectrum can be shaped by a diffraction grating. Separable Schmidt frequency modes can be also obtained in cavity-enhanced SPDC with sufficiently broadband pump and type-I phasematching. Our results, demonstrating the possibility of achieving a set of non-overlapping Schmidt modes forming a non-separable state, are expected to have a high impact on quantum state engineering.

{\bf Acknowledgements}
\par
The research leading to these results has received funding from the EU FP7 under grant agreement No.
308803 (project BRISQ2) and on the basis of Decision No. 912/2009/EC (project IND06-MIQC), NATO SPS Project 984397 ``SfP-Secure Quantum Communications'', Compagnia di S.~Paolo and MIUR (FIRB ``LiCHIS'' - RBFR10YQ3H, Progetto Premiale ``Oltre i limiti classici di misura''). We are grateful to the anonymous referee for pointing out the possibility to realize separable Schmidt modes with cavity SPDC.

\clearpage
\section{Supplemental information: Separable Schmidt modes of a non-separable state}





Here we provide additional information on the theory and experiment presented in the main paper. The
first part contains the calculation of Schmidt modes for a two-photon spectral amplitude (TPSA) given by a set of Gaussian functions. The second part analyzes the shape of a single maximum.
\maketitle

\vspace{5mm}

\subsection{Schmidt modes of a multi-peak TPSA.}

Consider the Schmidt modes for a TPSA given by a sum of $M$ double-Gaussian peaks,
\begin{equation}
F(x,y)=\sum_{i=1}^MA_i\exp\{-\frac{(x-x_{i})^2}{2\sigma^2}\}\,
\exp\{-\frac{(y-y_{i})^2}{2\sigma^2}\}, \label{Sum}
\end{equation}
where we denoted the signal and idler frequencies by $x$ and $y$ and assumed for simplicity that all Gaussians have the same width.

The case where the peaks overlap in both $x$ and $y$ is not interesting for the current work as it does not allow the lossless filtering of a single peak. Therefore, assume that the peaks do not overlap in $x$, i.e.,
\begin{equation}
\exp\{-\frac{(x-x_{i})^2}{2\sigma^2}\}\,\exp\{-\frac{(x-x_{j})^2}{2\sigma^2}\}=\delta_{ij}\exp\{-\frac{(x-x_{i})^2}{\sigma^2}\}, \label{nonoverlap}
\end{equation}
but partially overlap in $y$, so that
\begin{eqnarray}
\int d y\exp\{-\frac{(y-y_{i})^2}{2\sigma^2}\}\,
\exp\{-\frac{(y-y_{j})^2}{2\sigma^2}\}=c_{ij}\nonumber\\
c_{ii}=\int d y\exp\{-\frac{(y-y_{i})^2}{\sigma^2}\}\equiv c.
 \label{overlap}
\end{eqnarray}

In order to find the Schmidt modes, we need to solve the integral equation with the kernel~\cite{Law}
\begin{equation}
K_1(x,x')=\int dy F(x,y)F^*(x',y), \label{kernel_x}
\end{equation}
which is found to be
\begin{equation}
K_1(x,x')=\sum_{i,j=1}^MA_iA_j^*c_{ij}\exp\{-\frac{(x-x_{i})^2}{2\sigma^2}\}\,
\exp\{-\frac{(x'-x_{j})^2}{2\sigma^2}\}. \label{kernel_sum_x}
\end{equation}

The integral equation for finding the Schmidt modes $f_n(x)$ of variable $x$ is~\cite{Law}
\begin{equation}
\int dx' K_1(x,x')f_n(x')=\lambda_nf_n(x). \label{inteq_x}
\end{equation}

If the Schmidt mode is searched as one of the orthogonal normalized Gaussian functions,
\begin{equation}
f_n(x)=\frac{1}{c}\exp\{-\frac{(x-x_{n})^2}{2\sigma^2}\}, \label{sol_x}
\end{equation}
then Eq.~(\ref{inteq_x}) results in the equation
\begin{equation}
cA_n^*\sum_{i=1}^M c_{in}A_i\exp\{-\frac{(x-x_{i})^2}{2\sigma^2}\}=\lambda_n \exp\{-\frac{(x-x_{n})^2}{2\sigma^2}\}, \label{impossible}
\end{equation}
which is impossible to satisfy in the general case.
Therefore, this situation does not provide separable Schmidt modes.

In the special case where the Gaussians do not overlap in $y$ as well, $c_{ij}=c\delta_{ij}$, and Eq.~(\ref{impossible}) is satisfied with the eigenvalue
\begin{equation}
\lambda_n=c^2|A_n|^2. \label{sep_Schmidt}
\end{equation}
This special case is at the focus of the present paper as it enables lossless selection of a single-mode state out of an entangled one.

\subsection{A single peak of TPSA.}
Assume first that the shape of a single TPSA peak in Fig.~1b of the main text is given by a double Gaussian function,
\begin{equation}
	\label{TPSA_1FP}
F(\omega_{s},\omega_{i}) = \exp\{-\frac{(\sin\alpha\omega_s-\cos\alpha\omega_i)^2}{2\sigma_c^2}-\frac{(\omega_s+\omega_i)^2}{2\sigma_p^2}\},
	\end{equation}
where $\sigma_c$ is related to the crystal length $L$, $\sigma_p$ is given by the width of a single peak in the pump spectrum, $2\sqrt{2\ln2}\,\sigma_p$ being the full width at half maximum (FWHM), and $\Delta\omega$ is the distance between the peaks. The cross-section of this double Gaussian function, say, at half-maximum height, will be an ellipse oriented horizontally or vertically if the terms with the products $\omega_s\omega_i$ disappear, i.e.,
 \begin{equation}	
 \label{condition}
\sin(2\alpha)=\sigma_c^2/\sigma_p^2.
	\end{equation}

This alone does not ensure that the peak corresponds to a single term of the Schmidt decomposition for the whole state. Another necessary condition is that different Gaussian functions overlap neither in $\omega_i$ nor in $\omega_s$. This imposes two additional requirements:
\begin{eqnarray}	
 \label{conditions}
\left(\frac{\cos^2\alpha}{\sigma_c^2}+\frac{1}{2\sigma_p^2}\right)^{-1/2}<<\Delta\omega\cos\alpha,\nonumber\\
\left(\frac{\sin^2\alpha}{\sigma_c^2}+\frac{1}{2\sigma_p^2}\right)^{-1/2}<<\Delta\omega\sin\alpha.
	\end{eqnarray}

In the most common case of $\alpha<45^\circ$, the second condition is stronger. It can be written in the form
\begin{equation}	
\frac{\sin^4\alpha}{\sigma_c^2}+\frac{\sin^2\alpha}{2\sigma_p^2}>>\frac{1}{\Delta\omega^2}
 \label{cond}
	\end{equation}
and can be satisfied if $\Delta\omega$ is large enough.

Finally, consider a single TPSA maximum with an account for our real experimental conditions. The TPSA has the form (2) of the main text, with the pump spectral
amplitude given by the Fabry-Perot transmission in the vicinity of a single maximum,
\begin{equation}	
F_p(\omega)=\frac{1-R}{1-R\exp\{-2i\omega d\cos\phi/c\}},
 \label{FP_max}
	\end{equation}
where $R$ is the reflectance of each mirror (the mirrors are assumed to be similar), $d$ the FP spacing, $\phi$ the tilt of the plates and $c$ the speed of light.

\begin{figure}[h!]
\includegraphics[width=0.47\columnwidth]{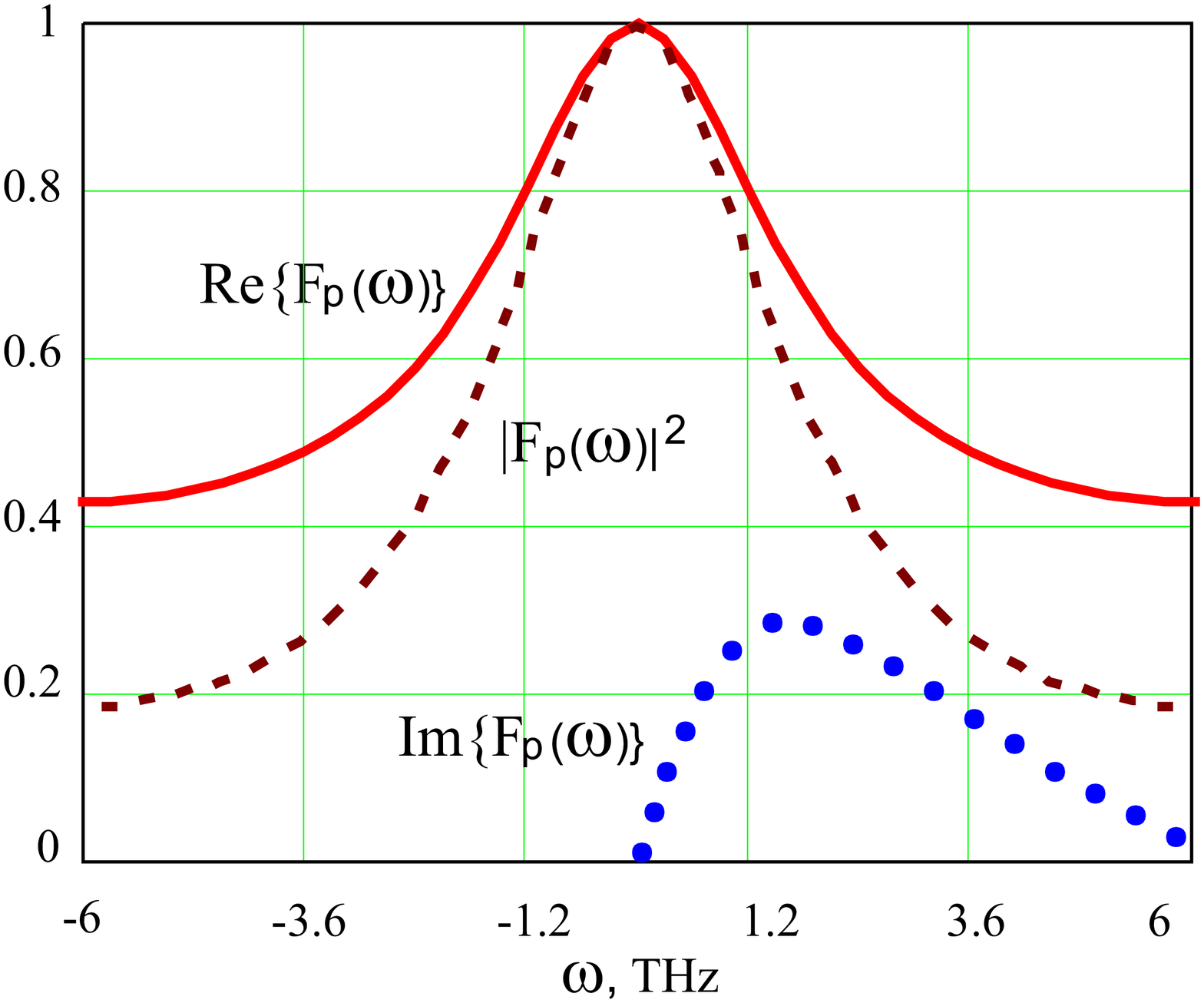}
\includegraphics[width=0.47\columnwidth]{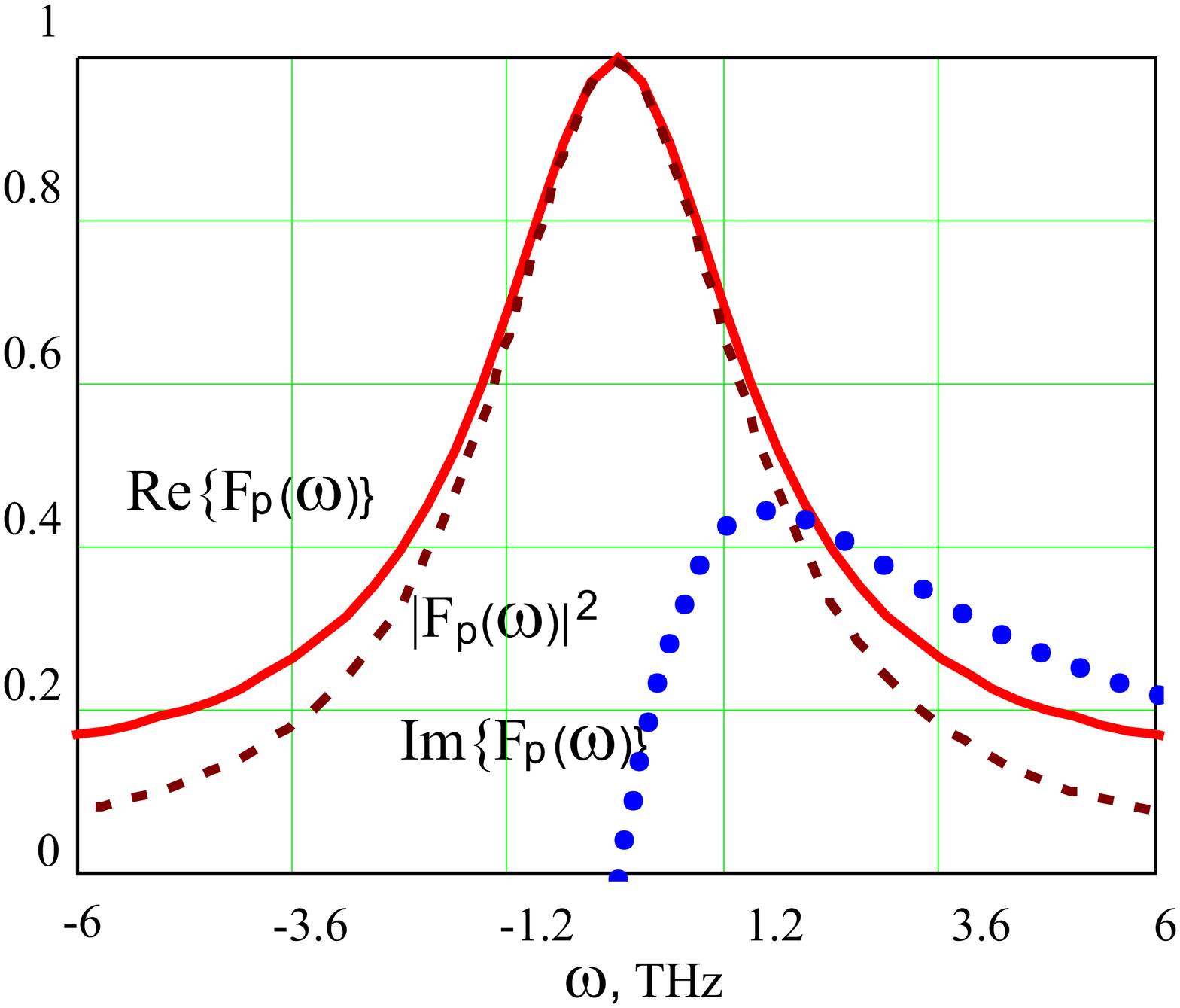}
\caption{The shape of the Fabry-Perot transmission $F_p(\omega)$ around a single peak. Red solid line: real part; blue dots: imaginary part (negative values not shown); brown dashed: squared modulus. Left: tilted FP, $d=100\,\mu$m, $\phi=45^\circ$, $R=0.4$. Right: untilted FP with the spacing changed to have the same width of a single maximum, $d=20\,\mu$m, $R=0.8$.} \label{pump}
\end{figure}
Figure~\ref{pump} shows the shape of $F_p$ including its real part (solid red), imaginary part (blue dots, with the antisymmetric negative values not shown), and the squared modulus (brown dashed) calculated with the parameters used in the experiment. Left panel shows the situation for FP tilted by $45^\circ$, described by $R=0.4,\,d=100\,\mu$m, which results in the wavelength FWHM $0.4$ nm near $\lambda=440$ nm. Right panel shows the case of untilted FP, giving the same pump width for $R=0.8,\,d=20\,\mu$m.

\begin{figure}[h!]
\vspace{10pt}
\includegraphics[width=0.45\columnwidth]{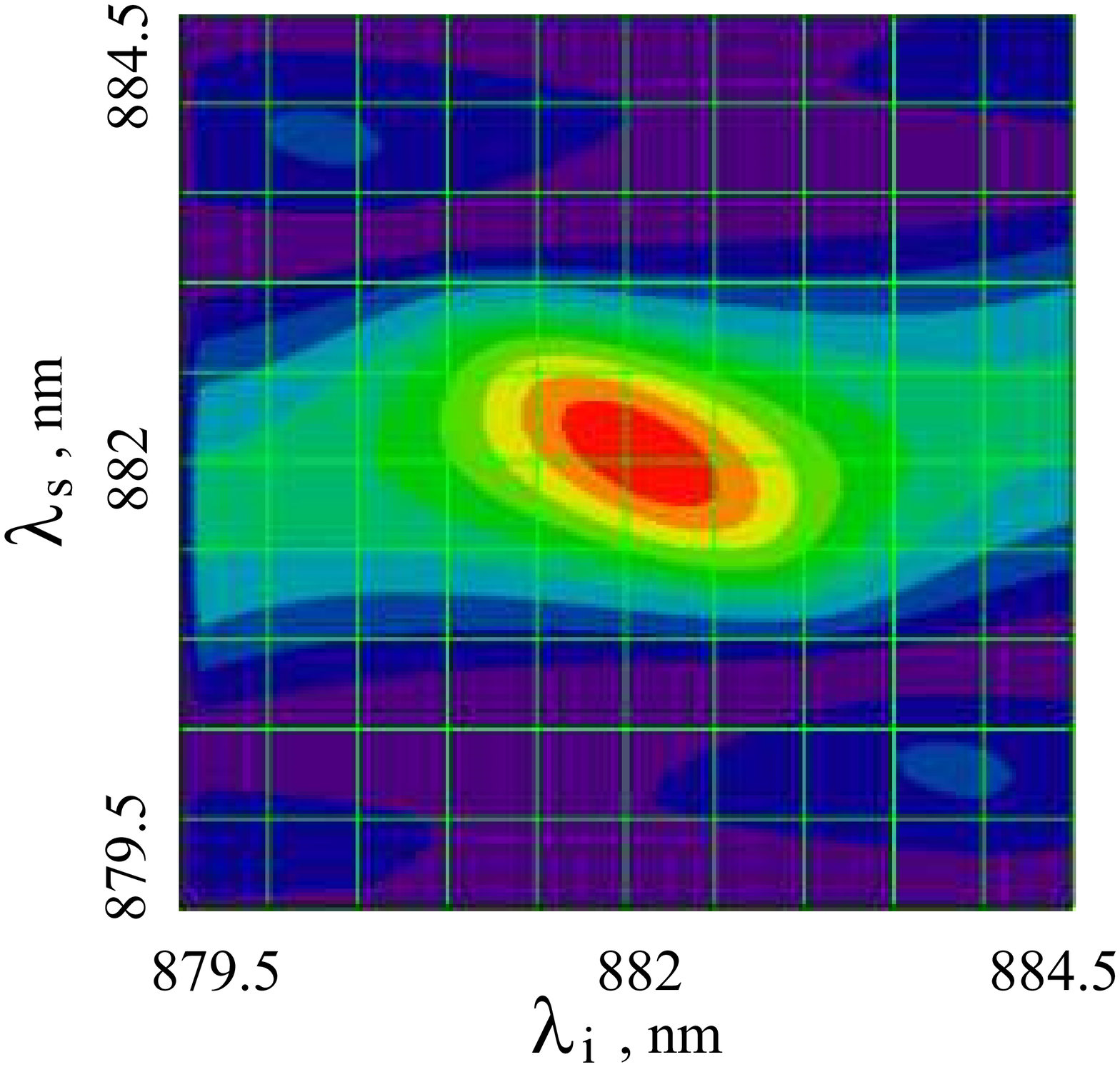}
\includegraphics[width=0.45\columnwidth]{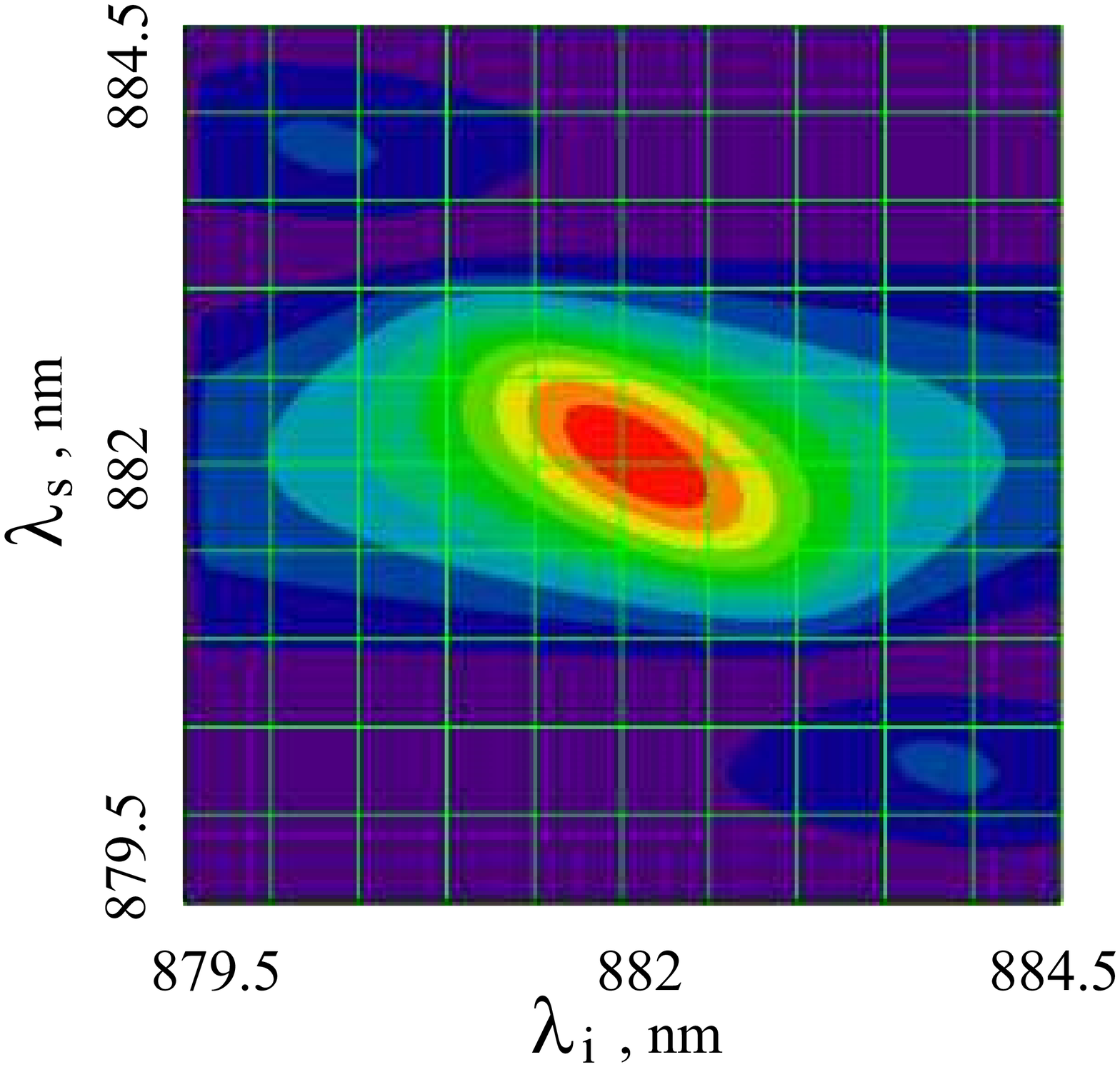}
\caption{The shape of the TPSA modulus around a single peak for the pump spectra shown in Fig.~\ref{pump}, pump central wavelength $440$ nm, and the crystal length 1.5 cm.} \label{TPSA}
\end{figure}
The TPSA for the corresponding cases is shown in Fig.~\ref{TPSA}. A lower finesse of a tilted FP results in a higher background.

It is worth noting that the TPSA has a frequency-dependent phase factor, caused both by the propagation of two-photon light through the crystal and the FP transmission function (\ref{FP_max}). This situation has been considered recently in Ref.~\cite{Gatti} and it was shown that if the dispersion can be neglected, it is only the modulus of TPSA that enters the Schmidt number calculation. In accordance with this, we have calculated the Schmidt number numerically using the modulus of the TPSA (Fig.~\ref{TPSA}). With cutting the spectrum exactly between the neighboring maxima (separated by 5 nm), we obtain for both cases $K=1.08$. This number is realistic for the higher-finesse case (right panel) where the spectrum is cut at $4\%$ of maximum intensity, and underestimated for the lower-finesse case (left panel), where the spectrum is cut at $16\%$ of the maximum.

\begin{figure}[h!]
\includegraphics[width=0.45\columnwidth]{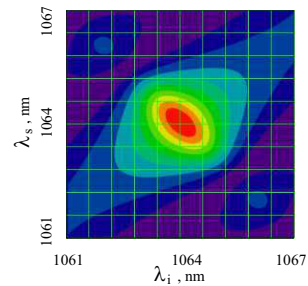}
\caption{The shape of the TPSA modulus around a single peak corresponding to the pump wavelength 532 nm, the FP spacing $d=50\,\mu$m, the FP transmission $R=0.64$, and the crystal length 20 mm.} \label{TPSA3}
\end{figure}

Finally, calculation of the TPSA for the case of pumping a $20$ mm crystal at $532$ nm with the 25 fs pump pulses passing through an untilted FP ($d=50\,\mu$m, $R=0.64$) gives $K=1.23$ with the intensity cut at $4\%$ of the maximum (Fig.~\ref{TPSA3}). Note that the smallest $K$ value is achieved when the `top' and the `wings' of the TPSA are tilted oppositely. In the case of the pump shaped as a comb of Gaussian peaks, the smallest $K$ is achieved at nearly circular TPSA maximum and amounts to $1.06$.

\end{document}